\documentclass[12pt]{article}

\usepackage[cp1251]{inputenc}  
\usepackage[T2A, OT1]{fontenc}
\usepackage[english]{babel}
\usepackage{amsfonts, longtable}
\usepackage{amsmath, amssymb, amsthm}
\usepackage{cite}
\usepackage{indentfirst}

\newtheorem*{theorem}{Theorem}
\newtheorem{lemma}{Lemma}
\newtheorem*{remark}{Remark}

\newcommand{\comment}[1]{}

\newenvironment{prf}{\par\textbf{Proof.}}{\par\qed\par}

\begin{document}

\title{Some Extensions of the Inversion Complexity \\ of Boolean Functions}
\author{V.\,V.\,Kochergin\footnote{ Lomonosov Moscow State University
 (Faculty of Mechanics and Mathematics,
 Bogoliubov Institute for Theoretical Problems of Microphysics); vvkoch@yandex.ru}, 
 A.\,V.\,Mikhailovich\footnote{National Research University Higher School of Economics; anna@mikhaylovich.com}}
\date{June, 2015}

\maketitle

\begin{abstract}
The minimum number of NOT gates in a Boolean circuit computing
a Boolean function is called the inversion complexity of the function.
In 1957, A. A. Markov determined the inversion complexity of every Boolean function
and proved that $\lceil\log_{2}(d(f)+1)\rceil$ NOT gates are necessary and sufficient to
compute any Boolean function $f$ (where $d(f)$ is maximum number of value changes from 1 to 0 over all increasing chains of tuples of variables values). In this paper we consider
Boolean circuits over an arbitrary basis that consists of all monotone functions (with zero weight) and finite nonempty set of non-monotone functions (with unit weight). It is shown that 
the minimal sufficient for a realization of the Boolean function $f$
number of non-monotone gates is equal to $\lceil\log_{2}(d(f)+1)\rceil - O(1)$.
Similar extends for another classical result of A. A. Markov for the inversion complexity of
system of Boolean functions has been obtained.

\medskip

{\it Keywords}: Boolean function complexity, circuit complexity, Boolean circuit complexity,
inversion complexity, Markov's theorem.
\end{abstract}

We denote the set $\{0,1\}$ by $E_2.$
A set of pairwise different tuples 
$$
\tilde{\alpha}_1=(\alpha_{11},\ldots, \alpha_{1n}),~
 \tilde{\alpha}_2=(\alpha_{21},\ldots, \alpha_{2n}),~ \ldots ,~
   \tilde{\alpha}_r=(\alpha_{r1},\ldots, \alpha_{rn})
$$
from the set $E_2^n$
such that 
$$
\alpha_{ij} \le \alpha_{i+1,j}, \quad i=1, \ldots, r-1,~~j=1, \ldots, n
$$
is called {\it increasing chain} (or {\it chain}).
The tuples $\tilde{\alpha}_1$ and $\tilde{\alpha}_r$ are called {\it initial} and
{\it terminal} tuples of the chain respectively.

Let $f(x_1,\ldots,x_n)$ be a Boolean function. 
An ordered pair of tuples $\tilde{\alpha}=(\alpha_{1},\ldots, \alpha_{n})$ and  $\tilde{\beta}=(\beta_{1},\ldots, \beta_{n})$, $\tilde{\alpha}, \tilde{\beta} \in E_k^n$
such that

1) ${\alpha_j} \le {\beta_j}$,~~$j=1, \ldots, n$;

2)  $f(\tilde{\alpha}) > f(\tilde{\beta})$

\noindent is called {\it jump}.
Lef $F=\{f_1(x_1,\ldots,x_n),\ldots,f_m(x_1,\ldots,x_n)\}$ be a set of Boolean functions.
A pair of tuples is called {\it jump for the system $F$} if 
there exists function $f\in F$ such that the pair is the jump for the function $f$.

Let $C$ be a chain of the form
$$
\tilde{\alpha}_1, \tilde{\alpha}_2, \ldots ,\tilde{\alpha}_r.
$$
The number of jumps for the system $F$ of the form $(\tilde{\alpha}_i, \tilde{\alpha}_{i+1})$
is called the {\it decrease $d_C(F)$ of the system $F$ along the chain} $C$.

The {\it decrease $d(F)$ of system $F$} is the maximum of $d_C(F)$ over all chains $C.$

\vspace{5mm}

We will investigate a realization of Boolean functions by circuits (or Boolean circuits, or logic circuit, or circuit of functional elements, or combinational machine~--- necessary definitions may be found in~\cite{LupMGU, Sav}) over basis $B$ of the form
$$
B= M \cup \{\omega_1, \ldots, \omega_p \}, \quad \omega_i \in P_2 \setminus M,~ i=1, \ldots, p, 
\eqno{(*)}
$$
and we assume that the weight of the functions from $M$ equals zero
and weight of the functions $\omega_1, \ldots, \omega_p$ equals 1. 

The complexity of circuit $S$ (function $f$, function system $F$ respectively) over basis $B$
of the form~$(*)$
is called {\it the inversion complexity over basis} $B$ and denoted by $I_B(S)$  ($I_B(f)$, $I_B(F)$ respectively).

Complete description of the inversion complexity of Boolean functions and 
systems of Boolean functions  over $B_0=M\cup \{ \overline{x} \}$ is given
by A.\,A.~Markov in~\cite{Mar57, Mar63}
(similar problems are also considered in~[5--16]): 
for any Boolean functions systems $F$ the equality
$$I_{B_0}(F) = \left\lceil \log_{2} (d(F)+1) \right\rceil $$
holds.

Since we can obtain negation (NOT gate) 
from any non-monotone function by substitution of constants, Markov's theorem implies 
the upper bound of the inversion complexity for arbitrary system of function over any basis $B$ of 
the form $(*)$:
$$I_{B}(F) \le \left\lceil \log_{2} (d(F)+1) \right\rceil .$$

If $B$ is an arbitrary basis such that for any $i$, $1\le i \le p,$ the equation $d(\omega_i)=1$ holds
then equality for inverion complexity from Markov's theorem holds (proof of lower bound
is similar to one in~\cite{Jukna}).

If a basis contains at least one function $\omega_j$ such that $d(\omega_j) > 1$, 
then equality does not hold. On the one hand, $I_{B}(\omega_j)=1$, on the other hand,
inequality $ \left\lceil \log_{2} (d(\omega_j)+1) \right\rceil \ge 2$ holds. 
Moreover, the inversion complexity of $F$ depends not only on $d(F)$ in that case.
Let us consider basis
$B_2= M  \cup \{ x \oplus y \oplus z \oplus 1 \}$ as an example. 
Let $F_1= \{x \oplus y \oplus z \oplus 1 \}$ and $F_2 = \{ \overline{x}, \overline{y} \}$.
Then
$$
d(F_1)=d(F_2)=2, \qquad  I_{B_2}(F_1)= 1,  \quad   I_{B_2}(F_2)= 2.
$$
To prove the last equality it is enough to note that for any $f(x,y)\in P_2$ inequality $d(f)\leq 1$ holds. 

The aim of this paper is to prove that 
despite of the fact mentioned above estimation of inversion complexity is almost the same 
for arbitrary basis $B$ of the form $(*)$ (even if $d(B)>1$). 
Let us give exact formulation of the result.
\begin{theorem} 
For any complete basis $B$ of the form $(*)$ there exists constant $c(B)$ such that for any system 
$F$ of Boolean functions
the following inequalities hold$:$
$$
 \left\lceil \log_2 (d(F)+1) \right\rceil  - c(B) \le   I_B(F)   \le \left\lceil \log_2 (d(F)+1) \right\rceil .
$$
\end{theorem}

We have already mentioned that the upper bound follows from Markov's theorem and 
the fact that negation can be obtained from any non-monotone function by substitution of constants.

Let us prove lower bound for arbitrary basis $B=M\cup \{\omega_1,\ldots,\omega_p\},$
where $\omega_i\in P_2\backslash M$ $(i=1,\ldots,p).$
Let $r(B)= \max \{ d(\omega_1), \ldots, d(\omega_p)  \}$. 

\begin{lemma}
Let $F\subset P_2.$ Then 
$$
d(F) \le (2r(B)+1) \left( 2^{I_B(F)} -1 \right).
$$
\end{lemma}
\begin{prf}
Let $F=\{f_1(x_1,\ldots,x_n),\ldots,f_m(x_1,\ldots,x_n)\}\subset P_2.$ 
The proof is by induction over $I_B(F)$.

If $I_B(F)=0$ then $F\subset M$. Hence, $d(F)=0$.

Assume the assertion is valid for any $G\subset P_2$ such that $I_B(G) \le I_B(F)-1.$ 

Let us consider a circuit $S$ with $n$ inputs $x_1,\ldots,x_n$ that realize system $F$ and 
contains exactly $I_B(F)$ vertices corresponding to functions from $\{\omega_1, \ldots, \omega_p \}$. 
Let us select the first such vertex (according to any correct numeration) and let us denote
corresponding gate by $E$. 
Denote by $h(x_1,\ldots,x_n)$ function that is obtained at the output of $E$. 
Denote by $S'$ a circuit that is obtained from the circuit $S$ by replacement the gate $E$ with one more input
with variable $y.$
The circuit $S'$ realize system
$G=\{g_1, \ldots , g_m \}$ with the following properties:
$$
f_i(x_1, \ldots , x_n)=
g_i\left(h(x_1, \ldots , x_n), x_1, \ldots , x_n \right), 
\quad i=1, \ldots, m.
$$
Moreover, $I_B(G) \le I_B(F)-1$.

Consider a chain C
$$
C=(\tilde{\alpha}_1, \tilde{\alpha}_2, \ldots ,\tilde{\alpha}_{r})
$$
such that $d(F)=d_C(F)$.

Let us consider the sequence $C'$ of $(n+1)$-tuples:
$$
({h(\tilde{\alpha}_1)},  \tilde{\alpha}_1), \ldots ,
({h(\tilde{\alpha}_r)}, \tilde{\alpha}_r).
$$
The sequence $C'$ is not a chain. 
Since 
$d(h) \le r(B)$, the first components of tuples from $C'$
change its values no more than $2r(B)+1$ times.
Denote by  $C_1'$ subsequence of $C'$ that includes all the tuples from $C'$ with first component equals zero 
and by $C_2'$ that includes all the tuples from $C'$ with first component equals one.
Both sequences $C_1'$ and $C_2'$ are chains of $n+1$-tuples.

By the inductive assumption inequalities
$$
d_{C_i'}(G) \le d(G) \le (2r(B)+1) \left( 2^{I_B(G)} -1  \right) = (2r(B)+1) \left( 2^{I_B(F)-1} -1  \right)
$$
hold, $i=1,2.$
Using equations 
$$
f_i (x_1, \ldots , x_n)=
g_i\left({h(x_1, \ldots , x_n)}, x_1, \ldots , x_n \right), 
\quad i=1, \ldots, m,
$$
we obtain
\begin{multline*}
d_C(F) \le  d_{C_1'}(G) + d_{C_2'}(G) + 2r(B)+1  \le \\
\le (2r(B)+1) \left( 2(2^{I_B(G)} -1) + 1  \right)
\le (2r(B)+1) \left( 2^{I_B(F)} -1 \right).
\end{multline*}

Thus, Lemma~1 is proved.
\end{prf}

Let $c(B)=\log_2(2r(B)+1) + 1.$

\begin{lemma}
Let $F\subset P_2.$ Then
$$
I_B(F) \ge \lceil \log_2 (d(F)+1) \rceil - c(B).
$$
\end{lemma}
\begin{prf}
Lemma~1 implies estimation 
$$
 2^{I_B(F)}  \ge \frac{d(F)}{2r(B)+1} +1.
$$
Hence,
\begin{multline*}
I_B(F) \ge \log_2 \left(  \frac{d(F)}{2r(B)+1} +1  \right) \ge \log_2  \frac{d(F)+1}{2r(B)+1} \ge \\
\ge \log_2 (d(F)+1) - \log_2 (2r(B)+1) \ge  \\
\ge \lceil \log_2 (d(F)+1) \rceil - \left( \log_2 (2r(B)+1) +1 \right)= \\
=  \lceil \log_2 (d(F)+1) \rceil - c(B).
\end{multline*}

Lemma~2 is proved.
\end{prf}

Lemma~2 completes the proof of the Theorem.

\begin{remark}
The result of Lemma 1 is rather rough. More accurately reasoning decreases value of c(B) in the Theorem.	
\end{remark}

\medskip 

This study (research grant No 14-01-0144) supported by The National Research University~--- Higher School of Economics' Academic Fund Program in 2014/2015.

The first author was supported by the Russian Foundation for Basic Research (project no.~14--01--00598).

\end{document}